\documentclass[epj]{svjour}

\usepackage{graphicx}
\usepackage{bm,bbold}
\usepackage{color,mathdots,tikz, xspace,amssymb}
\usepackage{hyperref}
\usepackage{subfigure}
\usepackage[normalem]{ulem}

\newcommand{\ket}[1]{{\ensuremath{|#1\rangle}}\xspace}
\newcommand{\bra}[1]{{\ensuremath{\langle #1|}}\xspace}
\newcommand{\diad}[1]{{\ensuremath{|#1\rangle\langle #1|}}\xspace}
\onecolumn

\begin{document}

\title{Dynamics of entanglement creation between two spins coupled to a chain}

%--------------------------------------
\author{Pierre Wendenbaum \inst{1}\inst{2} \and Bruno G. Taketani \inst{2}\inst{3} \and Endre Kajari\inst{2} \and Giovanna Morigi \inst{2} \and Dragi Karevski \inst{4}}
\institute{Institut Jean Lamour, dpt P2M, Groupe de Physique Statistique, Universit\'e de Lorraine-CNRS, B.P. 70239, F-54506 Vand\oe{}uvre-l\`es-Nancy Cedex, France.
\and Theoretische Physik, Universit\"at des Saarlandes, D-66123 Saarbr\"ucken, Germany.
\and Departamento de F\'isica, Universidade Federal de Santa Catarina, 88040-900, Florian\'opolis, Brazil.
\and Laboratoire de Physique et Chimie Th\'eoriques, UMR CNRS 7019, Universit\'e de Lorraine, BP 239 F-54506 Vand\oe{}uvre-l\`es-Nancy Cedex, France.}

\date{}

\abstract{We study the dynamics of entanglement between two spins which is created by the coupling to a common thermal reservoir. The reservoir is a spin-$\frac{1}{2}$ Ising transverse field chain thermally excited, the two defect spins couple to two spins of the chain which can be at a macroscopic distance.  In the weak-coupling and low-temperature limit the spin chain is mapped onto a bath of linearly interacting oscillators using the Holstein-Primakoff transformation. We analyse the time evolution of the density matrix of the two defect spins for transient times and deduce the entanglement which is generated by the common reservoir. We discuss several scenarios for different initial states of the two spins and for varying distances.
}
\maketitle

\section{Introduction}
Motivated by the fast advancements on the control of single quantum systems, the development of technologies for quantum communication \cite{Orieux2016,Yin2017,Ren2017}, quantum computation \cite{Takita2017,Neill2018} and quantum metrology \cite{Giovannetti2011} has bloomed in the past years. The key resource allowing quantum protocols to outperform classical ones is entanglement, which describes non-local correlations between different (distant \cite{Einstein1935}) physical objects and has no classical counterpart \cite{Horodecki2009,Feiler2013}. Nevertheless, due to its quantum nature, entanglement is very fragile against decoherence processes arising from interaction of the systems to their surrounding environment \cite{Zurek:2003vn,Schleich2000}.  Understanding the dynamics of entanglement in open systems is therefore of central importance. General results have been obtained in Refs. \cite{Benatti2003,Carvalho2007a,Mintert2005,Yu2002a,Hackermuller2004a,Konrad2007,Tiersch2013,Wendenbaum2014}. Moreover, many methods have been developed to protect entanglement against noise. The established strategies include error correction protocols \cite{Preskill1998}, quantum Zeno effect \cite{Facchi2004a,Maniscalco2008a} and weak measurements and measurement reversal \cite{Kim2011a}. 

A very different ansatz is based on quantum reservoir engineering (QRE). Here, by means of tailored interactions, the stochastic dynamics emerging from the interaction with an environment can drive a system into non-classical states \cite{Pielawa2007,Diehl2008,Verstraete2009,Pielawa2010a,Krauter2011a,Muschik2011}. This ideas have been successfuly applied, for instance, to create entangled atomic ensenbles \cite{Krauter2011a}, perform quantum simulations with trapped ions \cite{Barreiro2011}, superconducting qubits \cite{Messinger2019,Wang2019} and are the basis for protocols for quantum networks \cite{Pastawski2011,Vollbrecht2011} and quantum metrology \cite{Goldstein2011}. Environment-induced entanglement creation is closely related to the symmetries of the system under consideration. This feature has been extensively studied for example in spin chains \cite{CamposVenuti2006,CamposVenuti2007,Giampaolo2010} since this class of models offers in many cases the possibility of an analytic treatment. In particular, it has been shown that long-distance (or quasi long-distance) entanglement can be created between the boundary spins of a dimerized Heisenberg chain, thanks to the special form of its ground state  \cite{CamposVenuti2006,Giampaolo2010}. These examples rely on ground-state properties of finite chains. For increasing chain sizes, ground-state cooling represents a major technological challenge as the gap between ground and first excited states decreases exponentially fast with the chain size \cite{Giampaolo2010}. Mixing of ground and excited states in turn destroys entanglement. 

In contrast to protocols relying on ground-state properties, thermal environments at cold, but finite temperatures present a more robust channel to mediate interaction at a distance. One important ingredient is that also at finite temperature  symmetries of the dynamics can be tailored to create decoherence-free subspaces (DFS) \cite{Lidar2003}. Recently it has been shown that local coupling of two mass impurities to a thermal chain of oscillators represents one instance where such subspaces foster the creation and protection of entanglement \cite{Wolf2011,Kajari2012a}. These ideas are at the basis of protocols for generating  long-distance, steady-state entanglement in ion chains \cite{Fogarty2013,Taketani2014}.

In this work, we apply these concepts to spin chains. Our starting point is the model of Ref. \cite{Braun2002}, where entanglement between two spins coupled to the same locus of a harmonic chain has been characterized. We extend the model by assuming that the two ``defect'' spin $1/2$ particles interact each with distant elements of a long Ising chain. We show that in the high magnetic field and low temperature regime the interactions with the spin defects do not appreciably perturb the chain's state and yet can generate entanglement between the two defect spins. We also analyse the dependence of the entanglement dynamics on the distance between the spins.

The paper is organized as follows. In sec. \ref{sec:model}  we present the model and discuss the approximations used. We then summarize the salient properties of the chain and of the interaction dynamics, which are relevant to the generation of entanglement between the defect spins. The time evolution of the two defect spins is determined in Sec. \ref{sec:dynamics} and the entanglement properties are discussed as a function of time and of the distance between the spins they couple to. The conclusions are drawn in Sec. \ref{sec:conclusion}, while the appendices provide details of the calculations in Sec. \ref{sec:dynamics}.

%%%%%%%%%%%%%%%%%%%%%%%%%%%%%%%%%%%%%%%%%%%%%%%
%%%%%%%%%%%%%%%%%%%%%%%%%%%%%%%%%%%%%%%%%%%%%%%
\section{The model}
\label{sec:model}
We consider a transverse field  Ising chain made of $2N$ spins $1/2$ with periodic boundary conditions.
The Hamiltonian in terms of the Pauli $\sigma$-matrices is given by
\begin{equation}
H_b=-J{\sum_{<ij>}}^{'} \sigma_i^x\sigma_{j}^x-{\sum_{i}}^{'}  \sigma_i^z\; ,
\end{equation}
where the ferromagnetic ($J>0$) interaction runs over the nearest-neighbors $<ij>$ only and where
the label $i=0$ has been excluded from the sums $\sum'$ for further convenience\footnote{For chain sizes much larger than the defect spins separation, chains with even or odd number of spins should lead to equivalent dynamics.}.  At zero temperature this bath presents a quantum phase transition for $J=1$.
Two external ``defect" spins, labeled $A$ and $B$, 
experience a level splitting along $z$ described by the Zeeman Hamiltonian: 
\begin{equation}
H_d=-h (\sigma_A^z+\sigma_B^z)\; 
\label{eq:H_d}
\end{equation}
where $h$ is proportional to the ratio between the magnetic moments of the defect spins and the ones of the spins in the chain. 
The defect spins are locally coupled to two distinct sites of the Ising chain through the interaction Hamiltonian
 \begin{equation}
H_i=-\gamma (\sigma_A^x\sigma_{-l}^x+\sigma_B^x\sigma_l^x)\,,
\label{eq:H_i}
\end{equation}
where $2l$ is the distance between the sites and $\gamma$ is a coupling constant\footnote{As the environment has periodic boundary conditions, the choice of coupling positions is only restrictive in that the number of particles between the defect spins must be even.}.
The unitary dynamics is generated by the (dimensionless) Hamiltonian $H=H_b + H_i +H_d$. The energy is here given in units of the magnetic field along $z$ which the Ising chain experiences. 

We assume that the chain is prepared in a thermal state, $\rho_b=\exp[-H_b/T]/Z$, with temperature $T$ (Boltzmann constant $k_B=1$) and where $Z=\textmd{Tr}[\exp(-H_b/T)]$ is the chain partition function.  In the following, the two defect spins are initially prepared in a separable state, described by the density matrix $\rho_{A}(0)\otimes \rho_B(0)$, with $\rho_{A}$ and $\rho_{B}$ the density matrix of the defect spin $A$ and $B$, respectively. We remark that the defect spins do not directly couple. Instead, they couple to different sites of the Ising chain, which in turn undergo a collective dynamics. Correlations between the defect spins can thus solely emerge from the collective dynamics of the Ising chain.

%%%%%%%%%%%%%%%%%%%%%%%%%%
%%%%%%%%%%%%%%%%%%%%%%%%%%
\subsection{Bosonization}
\label{sec:bosonization}
As illustrated in Refs.\cite{CamposVenuti2006,CamposVenuti2007c,Giampaolo2010,Stauber2009a}, bath-mediated entanglement between the boundary spins of a finite spin chain is found in the ground-state of highly symmetric Hamiltonians where the reservoir presents long-range correlations. A major limitation faced in these systems is the vanishing energy gap between the ground state and the excited ones \cite{Giampaolo2010} as the chain size is increased. End-to-end entanglement is therefore very sensitive to decay of long-range correlations due to thermal fluctuations. In the case we consider here, decreasing the correlation length in the chain will decrease the correlations between spins $A$ and $B$. This will naturally lead to a drastic attenuation of the entanglement between the defects. 
With this in mind, we consider the paramagnetic regime, $J\ll1$, where in the reservoir's ground-state the spins are aligned with the transverse field. We perform then a bosonization of the bath spins by means of the Holstein-Primakoff transformation \cite{Holstein1940}, mapping spin modes into bosonic modes as
\begin{eqnarray}
&\sigma_n^+=\sqrt{1-a_n^\dagger a_n}\ a_n, \label{eq:HPT_lowering}\\
&\sigma_n^-=a_n^\dagger\sqrt{1-a_n^\dagger a_n},  \label{eq:HPT_raising}\\
&\sigma_n^z=\frac{1}{2}-a_n^\dagger a_n,  \label{eq:HPT_z}
\end{eqnarray}
where $\sigma^+_n$ and $\sigma^-_n$ are the raising and lowering spin operators and $a_n^{\dagger}$ and $a_n$ are the creation and annihilation operators of the bosonic mode $n$, satisfying the usual canonical bosonic commutation relations $[a_n,a_m^{\dagger}]=\delta_{n,m}$. Note that the expectation value of the bosonic number operator, $\langle a_n^\dagger a_n\rangle$, describes the deviation of the $n$-th spin magnetization from it's maximal value. In the weak inter-particle coupling and low temperature regime each reservoir spin is strongly polarized in the $z$-direction and Eqs.\ref{eq:HPT_lowering}-\ref{eq:HPT_z} can be expanded keeping only the zero-th order terms in the number operator  leading to
\begin{eqnarray}
&\sigma_n^+\approx a_n,  \\
&\sigma_n^-\approx a_n^\dagger.
\end{eqnarray}
For each bosonic mode $n$ we introduce the associated position and momentum operators, respectively $x_n$ and $p_n$,  defined by
\begin{eqnarray}
&a_n=\frac{1}{\sqrt{2}}(x_n+ip_n),\\
&a_n^\dagger=\frac{1}{\sqrt{2}}(x_n-ip_n)\; ,
\end{eqnarray}
leading to the canonical commutation rule $[x_n,p_m]=i\delta_{m,n}$ (we take $\hbar=1$).
In terms of these new operators  the total Hamiltonian $H$ takes the form
\begin{eqnarray}
H&=-2J\sum_n' x_nx_{n+1}+\frac{1}{2}\sum_n' (x_n^2+p_n^2) \nonumber \\
&-\sqrt{2}\gamma(x_{-l}\sigma_A^x+x_{l}\sigma_B^x)-h(\sigma_A^z+\sigma_B^z),
\label{eq:H_final}
\end{eqnarray}
where we have dropped an irrelevant constant term. 
Consequently the initial Ising chain has been mapped into a set of linearly coupled harmonic oscillators, 
with a minimal coupling to the spin defects via the local position operators $x_{-l}$ and $x_{l}$.

%%%%%%%%%%%%%%%%%%%%%%%%%%%%%%%%%%%%%%%%%%%%%%%
%%%%%%%%%%%%%%%%%%%%%%%%%%%%%%%%%%%%%%%%%%%%%%%
\subsection{Characterization of the bath}
\label{sec:charac_bath}
In order to determine the defects entanglement dynamics we first  diagonalize the bath Hamiltonian $H_b$. To shorten the notations we introduce position and momentum vectors, $\mathbf{x}^\dagger=(x_{-N},\cdots,x_{-1},x_1,\cdots,x_{N})$ and $\mathbf{p}^\dagger=(p_{-N},\cdots,p_{-1},p_1,\cdots,p_{N})$, such that
\begin{eqnarray}
H_b&=-2J\sum_nx_nx_{n+1}+\frac{1}{2}\sum_n(x_n^2+p_n^2) \nonumber \\
&=\frac{1}{2}\mathbf{p}^\dagger\mathbf{p}+\frac{1}{2}\mathbf{x}^\dagger  V_b \mathbf{x}\,,%= \frac{1}{2}{p}^2+\frac{1}{2}\mathbf{x}^\dagger V_b\mathbf{x},
\label{eq:H_bath}
\end{eqnarray}
with the potential matrix $V_b \in \mathbb{R}^{2N\times2N}$ defined as
\begin{equation}
V_b=2
\left(\begin{array}{ccccccc}
\displaystyle \frac{1}{2} &\displaystyle  -J & 0 & \cdots & \cdots & 0 &\displaystyle -J\\
\displaystyle -J & \ddots & \ddots & & & & 0  \\
0 & \ddots & \ddots & \ddots & & &\vdots  \\
\vdots & & \ddots & \ddots & \ddots & & \vdots \\
\vdots & & &  \ddots & \ddots & \ddots & 0 \\
0 & & & & \ddots & \ddots & \displaystyle -J  \\
\displaystyle -J & 0&\cdots &\cdots &0 &\displaystyle -J &\displaystyle \frac{1}{2}
\end{array}\right)
\,.
\end{equation}
The bath Hamiltonian, eq.(\ref{eq:H_bath}) (and more important, the coupling $H_i$) is invariant under the exchange of the $-n$ and $n$ bosons. It is thus advantageous to describe it in terms of symmetric (center-of-mass) and antisymmetric (relative) coordinates, defined as
\begin{equation}
x_n^{S,A}=\frac{x_n \pm x_{-n}}{\sqrt{2}}, \qquad  p_n^{S,A}=\frac{p_n \pm p_{-n}}{\sqrt{2}}
\label{eq:com_rel}
\end{equation}
where the superscripts $S$ and $A$ refers to center-of-mass and relative coordinates, respectively. In vectorial form, these relations read $\mathbf{\xi}=R\mathbf{x}$ and $\mathbf{\pi}=R\mathbf{p}$ with ${\mathbf{\xi}}^\dagger=({\mathbf{x}^{S}}^\dagger, {\mathbf{x}^{A}}^\dagger)=(x_1^{S},\dots, x_N^{S},x_1^{A},\dots, x_N^{A})$, ${\mathbf{\pi}}^\dagger =({\mathbf{p}^{S}}^\dagger, {\mathbf{p}^{A}}^\dagger)=(p_1^{S}, \dots, p_N^{S},p_1^{A}, \dots, p_N^{A} ) \in \mathbb{R}^{2N}$, and the orthogonal transformation matrix $R \in \mathbb{R}^{2N \times 2N}$
\begin{equation}
R=\frac{1}{\sqrt{2}}
\left(\begin{array}{cc}
\bar{\mathbb{1}} &\mathbb{1} \\
-\bar{\mathbb{1}} & \mathbb{1}
\end{array}\right)
,\textrm{with} \quad 
\bar{\mathbb{1}}=
\left(\begin{array}{ccc}
& & 1 \\
& \iddots & \\
1 &
\end{array}\right)
\in \mathbb{R}^{N \times N}.
\end{equation}
The transformed potential matrix, $\Lambda_b=RV_b R^\dagger$, assumes the block diagonal form
\begin{equation}
\Lambda_b=
\left(\begin{array}{cc}
V_b^S & 0 \\
0& V_b^A
\end{array}\right)
\label{eq:V_pm}
\end{equation}
with $V_b^S,V_b^A \in \mathbb{R}^{N\times N}$ the potential matrices for the center-of-mass (COM) and relative modes respectively. With this transformation it becomes evident that the dynamics occurs in two independent subspaces, $H_b=H_b^S + H_b^A$, with $H^{S,A}_b=\frac{1}{2} ({{p}^{S,A}})^2+\frac{1}{2}{{\mathbf{x}}^{S,A}}^\dagger V_b^{S,A}{\mathbf{x}}^{S,A} $.

Introducing the normal coordinates $\widetilde{\mathbf{x}}^{S,A}=({O^{S,A}})^\dagger\mathbf{x}^{S,A}$ and $\widetilde{\mathbf{p}}^{S,A}=({O^{S,A}})^\dagger\mathbf{p}^{S,A}$ where $O^{S,A}$ diagonalises the potential matrix $V_b^{S,A}$, $D^{S,A}=({O^{S,A}})^\dagger V_b^{S,A}  O^{S,A}$, the COM and relative Hamiltonians are given by a set of independent oscillators:
\begin{eqnarray}
\widetilde H^{S,A}_b=&\frac{1}{2} ({{\widetilde{p}}^{S,A}})^2+\frac{1}{2}(\widetilde{\mathbf{x}}^{S,A})^\dagger D^{S,A}\widetilde{\mathbf{x}}^{S,A} \nonumber \\
=&\sum_{n=1}^{N}\left(\frac{1}{2}(\widetilde{p}_n^{S,A})^ 2+\frac{1}{2}(\widetilde{\omega}_n^{S,A})^2\; (\widetilde{x}_n^{S,A})^2\right)
\end{eqnarray}
with eigenfrequencies 
\begin{eqnarray}
(\widetilde\omega_n^{S})^2=&1-4J\cos\left(\frac{n\pi}{N}\right), \quad n=0,\cdots,N-1 \label{eq:spectrum_COM}\\
(\widetilde\omega_n^{A})^2=&1-4J\cos\left(\frac{n\pi}{N}\right), \quad n=1,\cdots,N  \; . \label{eq:spectrum_rel}
\end{eqnarray}

%%%%%%%%%%%%%%%%%%%%%%%%%%
%%%%%%%%%%%%%%%%%%%%%%%%%%
\subsection{Hamiltonian in normal coordinates}
We now write the full Hamiltonian in the normal coordinates of the bath . In this way we can obtain the time-evolved density matrix of the spin defects. We shall first express the defect spins operator using  a non-local, symmetric and antisymmetric basis states:
$\ket{\phi^{S,A}}=(\ket{+,+}\pm\ket{-,-})/\sqrt{2}$ and $\ket{\psi^{S,A}}=(\ket{+,-}\pm\ket{-,+})/\sqrt{2}$  which are the (maximally) entangled Bell states and $\ket{\pm,\pm}$ the eigenstates of the $\sigma^z$ operators with eigenvalue $\pm1$.
Defining the $2$-dimensional operators  $S_x^S=(\ket{\psi^S}\bra{\phi^S}+\ket{\phi^S}\bra{\psi^S})$ and $S_x^A=(\ket{\psi^A}\bra{\phi^A}+\ket{\phi^A}\bra{\psi^A})$ which act as flip operators respectively on the symmetric and antisymmetric sectors of the defects Hilbert space, the 
 interaction Hamiltonian is rewritten as
\begin{eqnarray}
H_i=-2\sqrt{2}\gamma\bigg(S_x^S x_l^S 
+ S_x^A x_l^A\bigg)\; .
\end{eqnarray}
Again, as with the reservoir's free Hamiltonian, the coupling term also presents a clear separation between symmetric and antisymmetric parts. Introducing the coupling vector $\boldsymbol{\gamma}^T=\left(0,\cdots,2\sqrt{2}{\gamma},\cdots,0\right)$, the interaction Hamiltonian is rewritten 
\begin{equation}
{H}_i=-(S_x^S\boldsymbol{\gamma}^T\mathbf{x}^S+S_x^A\boldsymbol{\gamma}^T\mathbf{x}^A)\; .
\label{interaction}
\end{equation}
In terms of the reservoirs' normal coordinates, the interaction Hamiltonian is given by $\widetilde{H}_i=\widetilde{H}_i^S+\widetilde{H}_i^A$ where $\widetilde{H}_i^{S,A}=-S_x^{S,A}( \widetilde{\boldsymbol{\gamma}}^{S,A} )^T\widetilde{\mathbf{x}}^{S,A}$ and the coupling constants given by  $\widetilde{\boldsymbol{\gamma}}^{S,A}=(O^{S,A})^T\boldsymbol{\gamma}$. 
This is the starting point of Ref. \cite{Braun2002}, where entanglement generation was investigated in the case of two defects coupled at the same point of the spin chain. 

Finally, the Zeeman Hamiltonian of the defect spins is given in the new basis by
\begin{eqnarray}
\widetilde H_d&=&-2h\left(\ket{\phi^A}\bra{\phi^S}+ \ket{\phi^S}\bra{\phi^A}\right) \nonumber \\
&=&-2hS_z,
\label{eq:H_d_normal}
\end{eqnarray}
with $S_z\equiv \ket{\phi^S}\bra{\phi^A}+ \ket{\phi^A}\bra{\phi^S}$. Note that in the composed singlet/triplet basis $|j,m_j\rangle$ the defect Hamiltonian reads $\widetilde H_d=-2h(\diad{1,1}-\diad{1,-1})$. 
The full Hamiltonian reads $\widetilde{H}=\widetilde{H}^S+\widetilde{H}^A+\widetilde H_d$ with
\begin{eqnarray}
\widetilde{H}^S=&\frac{1}{2}(\widetilde{\mathbf{p}}^{S})^2+\frac{1}{2}(\widetilde{\mathbf{x}}^{S})^{\dagger}D^S\widetilde{\mathbf{x}}^{S}-S_x^S(\widetilde{\boldsymbol{\gamma}}^{S})^T\widetilde{\mathbf{x}}^S ,\\
\widetilde{H}^A=&\frac{1}{2}(\widetilde{\mathbf{p}}^{A})^2+\frac{1}{2}(\widetilde{\mathbf{x}}^{A})^{\dagger}D^A\widetilde{\mathbf{x}}^{A}-S_x^A(\widetilde{\boldsymbol{\gamma}}^{A})^T\widetilde{\mathbf{x}}^A.
\end{eqnarray}
While $\widetilde{H}^S$ and $\widetilde{H}^A$ describe independent dynamics for the symmetric and antisymmetric coordinates, \sout{and} the Zeeman part (\ref{eq:H_d_normal}) breaks this independance since $[\widetilde H_d,\widetilde H^{S,A}]\ne 0$.

%%%%%%%%%%%%%%%%%%%%%%%%%%%%%%%%%%%%%%%%%%%%%%%
%%%%%%%%%%%%%%%%%%%%%%%%%%%%%%%%%%%%%%%%%%%%%%%
\section{Dynamical generation of entanglement via the chain}
\label{sec:dynamics}

\subsection{Time evolution of the defects}
\label{sec:time_evo}

The non-commuting character of $\widetilde H_d$  with $\widetilde H^{S,A}$ renders the determination of the time evolution of the defects non trivial. Nevertheless, for time scales much shorter than the typical time set by the inverse defects energy gap $1/h$, the  contribution of the Hamiltonian $\widetilde H_d$ can be ignored at the lowest order in $h$ and a formal decoupled  solution  $\widetilde U(t)=\widetilde U^S(t)\widetilde U^A(t)$, with $\widetilde U^{S,A}(t)=e^{-it\widetilde H^{S,A}}$, is easily obtained. The time evolved reduced density matrix of the defects, $\rho_d(t)$, is then simply given by $\rho_d(t)=\textmd{Tr}_b\left[\widetilde U(t)\rho_{tot}(0)\widetilde U^\dagger(t)\right]$ where $\textmd{Tr}_b$ is the partial trace over chain's degrees of freedom and we assume that initially the system has been prepared in an uncorrelated product state
\begin{equation}
\rho_{tot}(0)=\rho_d(0) \otimes \rho_b(0)\; ,
\end{equation}
with a thermal environment $\rho_b(0)=\exp{(-\beta \widetilde H_b)}/Z$ and separable density matrix for the defect spins $\rho_d(0)=\rho_A(0) \otimes \rho_B(0)$. 
Within the above approximation, the matrix elements of the reduced density matrix $\rho_d(t)$ in the eigenbasis $\{\ket{s_i^{S,A}}\}$ of the operators $S_x^{S,A}$ ($S_x^{S,A}\ket{s_i^{S,A}}=s_i^{S,A}\ket{s_i^{S,A}}$ with eigenvalues $s_i^{S,A} =\pm 1$) are found to be (see appendix  \ref{app:Time_evolution} for details) 
\begin{eqnarray}
\bra{s_i}\rho_d(t)&\ket{s_j}\!=\!\exp\!
\left\{\!-\!\big[f^S(t)(s_i^{S}\!-\!s_j^{S})^2\!+\!f^A(t)(s_i^{A}\!-\!s_j^{A})^2\big]\right. \nonumber \\
&\left.+i\big[\varphi^S(t)((s_i^{S})^2\!-\!(s_j^{S})^2)\!+\!\varphi^A(t)(s_i^{A})^2\!-\!(s_j^{A})^2)\big]\right\} \nonumber \\
&\times \bra{s_i}\rho_d(0)\ket{s_j} 
,
\label{eq:elements_rho}
\end{eqnarray}
with the time dependent coefficients $f^{S,A}(t)$ and $\varphi^{S,A}(t)$ given by
\begin{eqnarray}
f^{S,A}(t)&=&\sum_{i} \frac{\left(\widetilde\gamma_i^{S,A}\right)^2(2\widetilde n^{S,A}_i-1)}{2\left(\widetilde\omega_i^{S,A}\right)^3}\left(1-\cos\left(\widetilde\omega_i^{S,A}t\right)\right),
\label{eq:f_pm}\\
\varphi^{S,A}(t)&=&\sum_{i} \frac{\left(\widetilde\gamma_i^{S,A}\right)^2}{2\left(\widetilde\omega_i^{S,A}\right)^2}\left(t-\frac{\sin\left(\widetilde\omega_i^{S,A}t\right)}{\widetilde\omega_i^{S,A}}\right).
\label{eq:varphi_pm}
\end{eqnarray}
Here, $\widetilde n^{S,A}_i$  is the thermal occupation of the mode $i$ of the symmetric (antisymmetric) chain given by $\widetilde n^{S,A}_i=\displaystyle (e^{\widetilde\omega_i^{S,A}/T}-1)^{-1}$. 
Note that every element appearing in the sum of Eq.(\ref{eq:f_pm}) is non-negative. The matrix elements of $\rho_d(t)$ have thus an oscillatory term, dependent on $\varphi^{S,A}(t)$, and an exponential time-dependent attenuation set by $f^{S,A}(t)$. In particular, we remark that the diagonal elements of the defect states do not evolve in time. The spin chain therefore constitutes a purely dephasing environment in the $\{\ket{s_i^{S,A}}\}$ basis.

%%%%%%%%%%%%%%%%%%%%%%%%%%%%%%%%%%%%%%%%%%%%%%%
%%%%%%%%%%%%%%%%%%%%%%%%%%%%%%%%%%%%%%%%%%%%%%%
\subsection{Entanglement dynamics}
\label{sec:ent_dynamics}
Entanglement is hereafter quantified by the Concurrence \cite{Wootters1998,Horodecki2009}, which for two qubits takes the form 
$$C(\rho(t))=\textrm{max}\{0, \lambda_1-\lambda_2-\lambda_3-\lambda_4\}\,.$$ Here, $\lambda_1,\ldots,\lambda_4$ are the square root of the eigenvalues in decreasing order of the (generally) non-hermitian matrix $\rho\tilde{\rho}$ with $\tilde{\rho}=(\sigma_y \otimes \sigma_y)\rho^*(\sigma_y \otimes \sigma_y)$, where the complex conjugate is taken in the standard basis. In the rest of this section we consider two perfect degenerated defect spins, {\it i.e.}, $h=0$, which allows us to determine the dynamics for long times.

%%%%%%%%%%%%%%%%%%%%%%%%%%
%%%%%%%%%%%%%%%%%%%%%%%%%%
\subsubsection{Spins coupled to the same site}
We first analyse the case where both spin defects couple to the same chain site $l$. This situation was first considered in Ref.\cite{Braun2002}. In this case the interaction Hamiltonian simplifies to
\begin{eqnarray}
\widetilde{H}_i=S_x(\boldsymbol{\widetilde\gamma}^S\widetilde{\mathbf{x}}^S+\boldsymbol{\widetilde\gamma}^A\widetilde{\mathbf{x}}^A)\; .
\end{eqnarray}
Note that the antisymmetric states $\ket{\psi^A}$ and $\ket{\phi^A}$ do not appear here. Coupling the spins at the same point of the chain thus creates a $2$-dimensional decoherence-free subspace \cite{Lidar2003}, protected from the non-unitary dynamics set by the bath.

\begin{figure}[tp]
\centering
\includegraphics[width=0.95\textwidth]{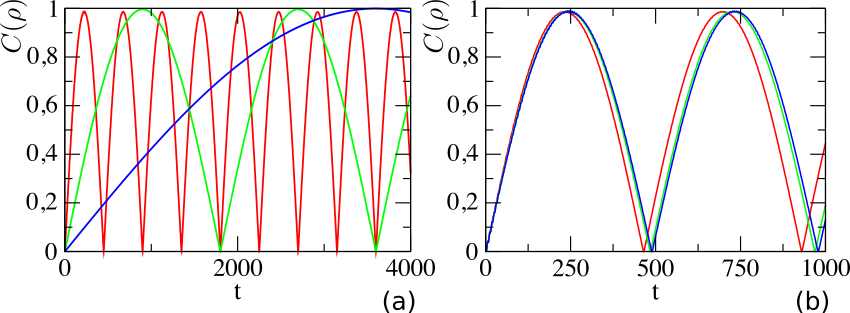}
\caption{Concurrence dynamics for defect spins coupled at the same point and with initial state $\ket{\varphi_A}\otimes \ket{\varphi_B}=\ket{\uparrow_A} \otimes \ket{\uparrow_B}$. Parameters are (a) $l=10$, $J=0.2$ $T=0.00001$ and $\gamma=0.04$ (red), $\gamma=0.02$ (green), $\gamma=0.01$ (blue) and (b) $l=10$, $\gamma=0.04$ $T=0.00001$ and $J=0.16$ (red), $J=0.08$ (green), $J=0.04$ (blue). 
}
\label{fig:samepoint}
\end{figure}

Figure \ref{fig:samepoint} shows the concurrence $C(t)$ calculated for the initial defects' state $\ket{\uparrow_A}\otimes\ket{\uparrow_B}$, namely, when both spin are parallel and aligned along $x$. The dynamics of the concurrence $C$ is calculated for different values of $\gamma$ at constant $J$ (a), and for different values of $J$ at given $\gamma$ (b).
We observe that entanglement oscillates with a period depending on the chain parameters and reaches a maximum value close to $1$.

\begin{figure}[h]
\centering
\includegraphics[width=0.65\textwidth]{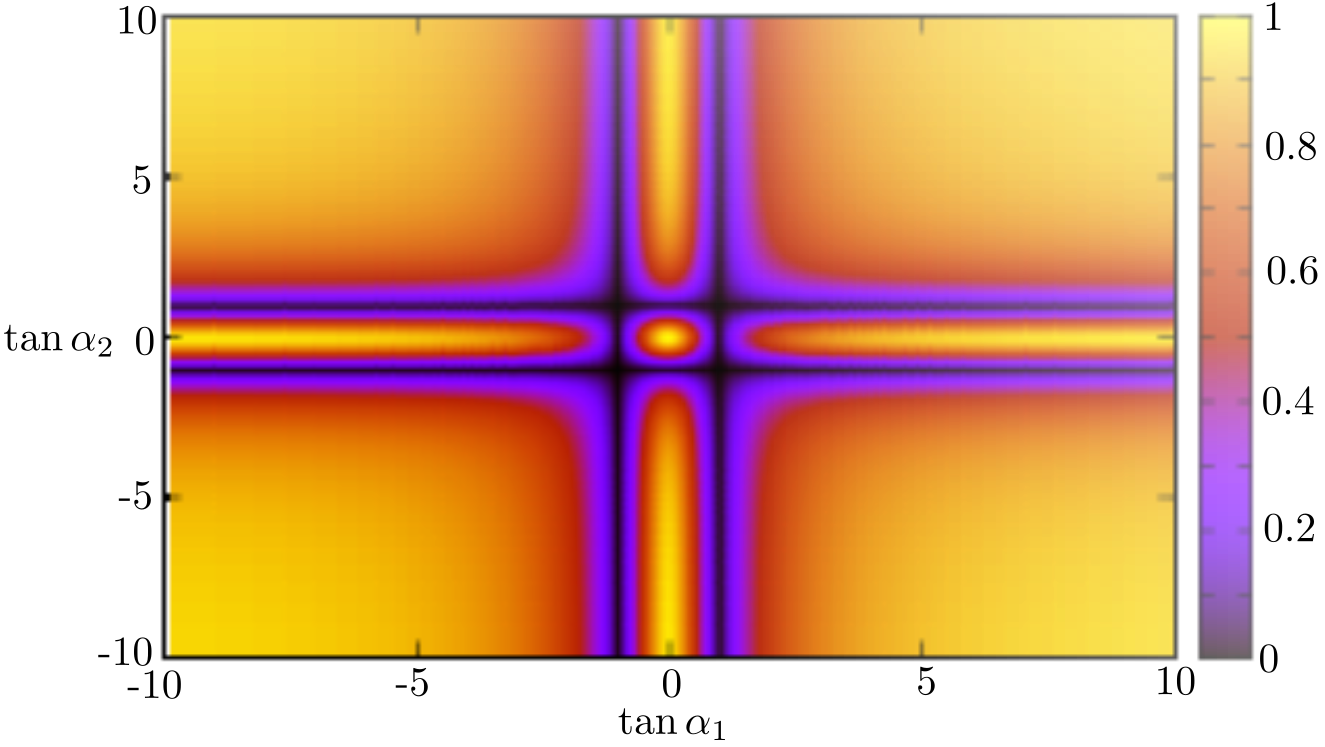}
\caption{Maximal concurrence, $\max_t C(t)$, as a function of the initial state of the defect spins, $\rho_{A,B}= \ket{\varphi_A}\bra{\varphi_B}$, with $\ket{\varphi}_i=\cos{\alpha_i}\ket{\uparrow}_i+\sin{\alpha_i}\ket{\downarrow}_i$. The parameters are $J=0.2$, $\gamma=0.04$ and $T=0.00001$, the spin defects are coupled at the same chain site. }
\label{fig:initial_state}
\end{figure}

We have further verified that  the amplitude and period of the observed oscillations depend on the initial state of the defects. Fig.\ref{fig:initial_state} displays the maximal value attained by the concurrence as a function of the initial state $\ket{\varphi_A}\otimes \ket{\varphi_B}$, with $\ket{\varphi}_i=\cos{\alpha_i}\ket{\uparrow}_i+\sin{\alpha_i}\ket{\downarrow}_i$, and $i=A,B$. Entanglement is found for any initial state, except when at least one of the spins is prepared in an eigenstate of $\sigma^x$. We understand the oscillations as precessions of the defects spin due to the magnetic field of the chain spin to which they couple. This explains the dependence of the angle $\alpha_i$ observed in Fig. \ref{fig:initial_state}.

%%%%%%%%%%%%%%%%%%%%%%%%%%
%%%%%%%%%%%%%%%%%%%%%%%%%%
\subsubsection{Entanglement between distant spins}
\begin{center}
\begin{figure}[hb]
\centering
\includegraphics[width=0.95\textwidth]{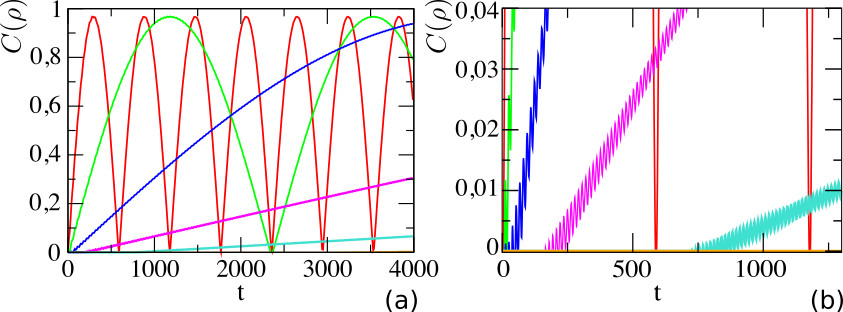}
\caption{Time evolution of the concurrence for $J=0.2$, $\gamma=0.04$, $T=0.00001$ and distances $l=1$ (red), $l=2$ (green), $l=3$ (blue), $l=4$ (pink), $l=5$ (light blue) and $l=6$ (orange). (b) Zoom of plot (a) showing the dynamics of  $C(t)$. The initial state is $\ket{\varphi_A}\otimes \ket{\varphi_B}=\ket
{\uparrow_A} \otimes \ket{\uparrow_B}$.}
\label{fig:distance}
\end{figure}
\end{center}

The decoherence-free subspace we identified in the previous subsection is a direct consequence of the symmetry of the Hamiltonian: when the defects spins couple to the same site, the triplet components of the defect spin couples to the chain, while the singlet (which is the antisymmetric component according to our formalism) is decoupled. The singlet state is a decoherence free subspace when $h=0$, namely, when the Zeeman Hamiltonian $\widetilde{H_d}$ vanishes. We consider now the case where the defects spins are prepared in a separable state and couple to two distant sites. Specifically, spin-$A$ is coupled to the $l$-th chain spin and spin-$B$ to the $-l$-th chain spin. Figure \ref{fig:distance} shows the dynamics of their concurrence for varying inter-particle distances $2l-1$, $J=0.2$ and $\gamma=0.04$. Also in this case entanglement undergoes oscillations in time and the  oscillation period decreases with increasing $\gamma$. However, one may notice two main differences: (i) the oscillation period increases with the defects distance and (ii) entanglement creation is not instantaneous, but rather takes a finite time $t_0$ to set in, as can be seen from Fig. \ref{fig:distance}(b). From the data of Fig. \ref{fig:distance}(a) we found that the onset time $t_0$ grows exponentially with the distance. Moreover, entanglement oscillates with frequencies that decrease exponentially with distance.

The exponential growth of $t_0$ with the distance separating the two defects cannot be explained by the time taken for information to travel through the chain, from one defect to the other: this would rather lead to a linear growth. This behaviour, instead, indicates that by changing the defects' positions the bath modes that mediates the interaction also changes. Moreover, in Fig. \ref{fig:distance}(b) the concurrence exhibits fast oscillations about the mean value. This indicates that the interaction is mediated my multiple modes, however, a single mode dominates and determines the periodicity of the concurrence.

%%%%%%%%%%%%%%%%%%%%%%%%%%%%%%%%%%%%%%%%%%%%%%%
%%%%%%%%%%%%%%%%%%%%%%%%%%%%%%%%%%%%%%%%%%%%%%%
\subsection{Entanglement and spectral density}
\label{sec:sd}
The entanglement dynamics discussed in the last section is a feature of the degeneracy of the triplet states of the two defect spins system $\ket{1,1}$ and $\ket{1,-1}$, when the free defect spins are degenerate in absence of the chain ($h=0$). One consequence of this assumption is the appearance of a decoherence-free subspace which promotes immediate entanglement generation and is protected by the environment due to symmetry. For the general case where $h\neq0$, we will now show that techniques previously studied by some of the authors to identify decoherence-free subspaces cannot be applied \cite{Wolf2011,Kajari2012a,Fogarty2013,Taketani2014}. In this case the decoherence-free subspace of $h=0$ disappears, since the magnetic field couples the symmetric and antisymmetric Hilbert subspaces of the defects. Nevertheless, the dynamics offer other hidden symmetries which emerge from the reflection symmetry about $l=0$. In this case, in fact, for certain values of $h$ the spin defects can be seen as forming two effective hard boundaries, supporting local modes of the chain involving the defects and the chain spins separating them. This condition has been identified and applied in Refs.\cite{Wolf2011,Kajari2012a,Fogarty2013,Taketani2014} for mass defects coupled to a chain of oscillators. There, it was shown that this dynamics can generate entanglement between the mass defects, even when these are coupled to distant sites of the chain. We now discuss the validity of applying these concepts to the spin defects, given the constraints imposed by the bosonization procedure. In order to identify the values to which $h$ shall be tuned, we follow the prescription of Ref. \cite{Taketani2014}. For this purpose we analyze the spectral-density functions \cite{Weiss1999} of the symmetric and antisymmetric chains, which are defined as: 
\begin{equation}
\label{eq:SpectralDensity}
\mathcal I_{S,A}(\widetilde{\omega})=\frac{\pi}{2}\sum_n\frac{(\widetilde\gamma_n^{S,A})^{2}}{\widetilde{\omega}_n^{S,A}}\delta\left(\widetilde{\omega}-\widetilde{\omega}_n^{S,A}\right)\,.
\end{equation}
\begin{center}
\begin{figure}[t]
\centering
\includegraphics[width=0.95\textwidth]{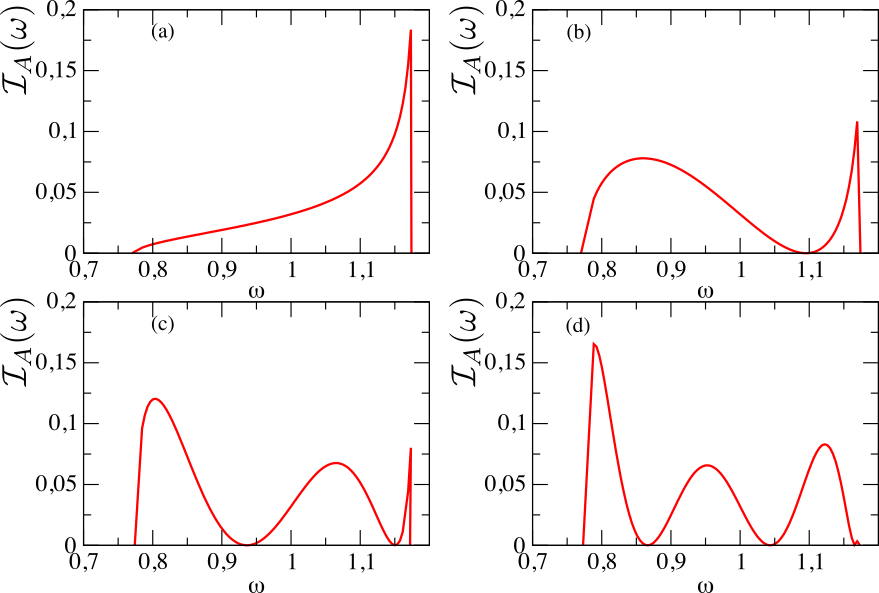}
\caption{Spectral density associated to the relative coordinates bath for several values of the distance. The parameters are $\gamma=0.04, T=0.0001$, $J=0.2$. The values of $l$ are $l=1$ (a), $l=2$ (b), $l=3$ (c), and $l=4$ (d).}
\label{fig:spectral_density}
\end{figure}
\end{center}
Figure \ref{fig:spectral_density} shows example spectral densities for different defects separations $2l-1$ and large $N$. The existence of zeros in $\mathcal{I}_{S,A}$ indicate the presence of the aforementioned decoherence-free subspaces. The nodes are obtained taking the continuum limit of Eq. \ref{eq:SpectralDensity}, which deliver the frequencies 
\begin{equation}
\widetilde{\omega}_0^A(p)=\sqrt{1-\displaystyle 4J\cos\left(\displaystyle \frac{2p\pi}{2l-1}\right)}\,,
\end{equation}
with $p=0,\cdots,l-1$ and $l-1$ non-trivial zeroes for a given distance $l$. The zeroes of the spectral densities occur for $\widetilde\omega\in [\sqrt{1-4J},\sqrt{1+4J}]$.

Following Refs.\cite{Wolf2011,Kajari2012a,Fogarty2013,Taketani2014}, steady-state entanglement couple be achieved by tuning the defect spins transition frequency $h$ to match one of the zeroes of $\mathcal{I}_{S,A}(\widetilde\omega)$. With, e.g., $h$ engineered to match a given $\widetilde{\omega}_0^A(p)$, the defects' non-local states $\ket{\phi^A}$ and $\ket{\psi^A}$ effectively decouples from the corresponding antisymmetric bath and form the decoherence-free subspace. However, the bosonization procedure of sec.\ref{sec:bosonization} imposes small intra-chain coupling strenght $J\ll 1$. Tuning the defect's transition frequencies to a zero of the spectral densities therefore requires $h\approx 1$, rendering the results of the previous section invalid. A direct application of the methods described in Ref.\cite{Wolf2011,Kajari2012a,Fogarty2013,Taketani2014} is therefore not possible within our model.

%%%%%%%%%%%%%%%%%%%%%%%%%%%%%%%%%%%%%%%%%%%%%%%
%%%%%%%%%%%%%%%%%%%%%%%%%%%%%%%%%%%%%%%%%%%%%%%
\section{Conclusion}
\label{sec:conclusion}
We have investigated the dynamics of entanglement generation between two distant spin $1/2$ particles coupled to a common spin reservoir. Within the low-temperature and weak coupling approximation, we have shown that entanglement can be generated for spins coupled locally to distant elements of the environment. As this distance increases, the time taken for onset of entanglement generation increases exponentially. This takes place as the bath modes mediating the interaction vary with the distance. For degenerate spins, we have shown that the generated entanglement varies periodically and that the period also depends on the distance. We have further explored the possibility of entanglement generation when the parameters are tuned so to generate decoherence free subspaces involving the excitation of the defects spins, following the protocol identified in Refs.\cite{Wolf2011,Kajari2012a,Fogarty2013,Taketani2014}. We found that these decoherence-free subspaces lie outside the parameter regime of the proposed model and steady-state entanglement generation cannot be achieved following these methods.

In this work we restricted to the regime in which the Ising chain is deep in the ordered phase. It is further interesting to analyse how these dynamics are modified when the chain is close to criticality \cite{Osterloh2002,Cucchietti2007,Baltrusch2012}, and thus to investigate the interplay between decoherence-free subspaces, emerging from the microscopic details, and critical behaviour. Localization due to disorder in the chain  \cite{Toermae2002} is a further resource for promoting entanglement creation and will be object of future work. 

\begin{acknowledgement}
The authors are grateful to Alexander Wolf and Thomas Fogarty for helpful discussions and to Wolfgang Schleich for his insightful advice and for inspiring and stimulating discussions. G.M. acknowledges support from DPG SPP 1929. B.G.T. also acknowledges support from FAPESC and CNPq INCT-IQ (465469/2014-0).
\end{acknowledgement}

%%%%%%%%%%%%%%%%%%%%%%%%%%%%%%%%%%%%%%%%%%%%%%%
%%%%%%%%%%%%%%%%%%%%%%%%%%%%%%%%%%%%%%%%%%%%%%
\appendix

\section{Time evolution of the reduced density matrix}
\label{app:Time_evolution}
Under the assumption of vanishing defects free Hamiltonian, the full dynamics is described by
\begin{eqnarray}
\widetilde{H}=&\frac{1}{2}(\widetilde{\mathbf{p}}^{S})^2+\frac{1}{2}(\widetilde{\mathbf{x}}^{S})^TD^S\widetilde{\mathbf{x}}^{S}+S_x^S(\widetilde\gamma^{S})^T\widetilde{\mathbf{x}}^S \nonumber \\
+&\frac{1}{2}(\widetilde{\mathbf{p}}^{A})^2+\frac{1}{2}(\widetilde{\mathbf{x}}^{A})^TD^A\widetilde{\mathbf{x}}^{A}+S_x^A(\widetilde\gamma^{A})^T\widetilde{\mathbf{x}}^A \nonumber \\
=& \widetilde{H}_b^S +\widetilde{H}_i^S + \widetilde{H}_b^A + \widetilde{H}_i^A.
\end{eqnarray}
In the interaction picture with respect to the bath Hamiltonian we find
\begin{equation}
\widetilde{H}^{(I)}=e^{i\widetilde{H}_b^St}\widetilde{H}_i^Se^{-i\widetilde{H}_b^St}+e^{i\widetilde{H}_b^At}\widetilde{H}_i^Ae^{-i\widetilde{H}_b^At}.
\end{equation}
Using Baker-Campbell-Hausdorff expansion one easily finds
\begin{eqnarray}
\widetilde{H}^{(I)}&=\sum_i\bigg(\gamma_i^S\cos(\widetilde{\omega}_i^St)S_x^S\widetilde x_i^S+\frac{\gamma_i^S}{\widetilde{\omega}_i^S}S_x^S\widetilde p_i^S \nonumber \\
&+\gamma_i^A\cos(\widetilde{\omega}_i^At)S_x^A\widetilde x_i^A+\frac{\gamma_i^A}{\widetilde{\omega}_i^A}S_x^A\widetilde p_i^A\bigg).
\label{Hamil}
\end{eqnarray}
Neglecting the free defects Hamiltonian permits us to write the full time evolution operator as $\widetilde  U(t)=\widetilde U^S(t)\widetilde U^A(t)$. Operators $\widetilde U^{(S,A)}(t)$ can now be obtained with the Ansatz 
\begin{eqnarray}
&\widetilde U^{S,A}(t)=\exp{\left(\displaystyle i\sum_i\delta_i^{S,A}(t)\right)} \nonumber \\
&\times\exp{\left(i\displaystyle\sum_i\big(\phi_i^{S,A}(t)\widetilde p_i^{S,A}+(\phi_i^{S,A})'(t)\widetilde x_i^{S,A}\big)\right)}.
\end{eqnarray}
Deriving $\widetilde U(t)$ with respect to time
\begin{eqnarray}
&\dot{\widetilde U}=i\sum_i\bigg( \dot{\delta}_i^S+\dot{\phi}_i^S\widetilde p_i^S-\frac{1}{2} \frac{d}{dt}\big(\phi_i^S(\phi_i^{S})'\big)\bigg)\widetilde{U}^S\widetilde{U}^A \nonumber \\
&+i\sum_i\bigg( \dot{\phi}_i^S\widetilde x_i^S+(\dot{\phi}_i^{S})'\phi_i^S\bigg)\widetilde{U}^S\widetilde{U}^A \nonumber \\
&+i\sum_i\bigg(\dot{\delta}^A_i+\dot{\phi}_i^A\widetilde p_i^A-\frac{1}{2}\frac{d}{d t}\big(\phi_i^A(\phi_i^{A})'\big)\bigg) \widetilde{U}^S\widetilde{U}^A \nonumber \\
&+i\sum_i\bigg( \dot{\phi}_i^A\widetilde x_i^A+(\dot{\phi}_i^{A})'\phi_i^A\bigg)\widetilde{U}^S\widetilde{U}^A,
\end{eqnarray}
where the dot represents time derivative and the explicit time dependence has been omitted. Schr\"odinger's equation for the evolution operator $\dot{\widetilde U}(t)=-i\widetilde{H}^{(I)}(t)\widetilde U(t)$ can be now used to find $\phi_i^{A(B)}$ and $\phi_i^{'A(B)}$. Using the initial condition $\widetilde U(0)=\mathbb{1}$, we find
\begin{eqnarray}
\phi_i^{S,A}( t)=&-\frac{\gamma_i^{S,A}S_x^{S,A}}{\widetilde{\omega}_i^{S,A}}\sin\left(\widetilde{\omega}_i^{S,A} t\right) \nonumber \\
(\phi_i^{S,A})'( t)=&-\frac{\gamma_i^{S,A}S_x^{S,A}}{\widetilde{\omega}_i^{S,A}}\left(\cos\left(\widetilde{\omega}_i^{S,A} t\right)-1\right) \nonumber \\
\delta_i^{S,A}( t)=&\frac{(\gamma_i^{S,A})^2(S_x^{S,A})^2}{\widetilde{\omega}_i^{S,A}}\bigg( t-\frac{\sin\left(\widetilde{\omega}_i^{S,A} t\right)}{ \widetilde{\omega}_i^{S,A}}\bigg).
\end{eqnarray}
The evolution operator is seen to be a displacement operator
\begin{equation}
\widetilde{U}( t)=e^{i \sum_i\big(\delta_i^S(t)+\delta_i^A(t)\big)}e^{i(\mathbf{Q}^S\mathbf{\widetilde x}^S-\mathbf{R}^S\mathbf{\widetilde p}^S)}e^{i(\mathbf{Q}^A\mathbf{\widetilde x}^A-\mathbf{R}^A\mathbf{\widetilde p}^A)},
\end{equation}
with
\begin{eqnarray}
&\mathbf{Q}^{S,A}=-\sum_i\frac{{\gamma}_i^{S,A}}{{\widetilde \omega}_i^{S,A}}\sin\left({\widetilde{\omega}}_i^{S,A} t\right)S_x^{S,A}\mathbf{e}_i^{S,A}, \nonumber \\
&\mathbf{R}^{S,A}=-\sum_i\frac{{\gamma}_i^{S,A}}{{\big(\widetilde\omega}_i^{S,A}\big)^2}\left(\cos\left({\widetilde{\omega}}_i^{S,A} t\right)-1\right)S_x^{S,A}\mathbf{e}_i^{S,A}
\end{eqnarray}
and $\mathbf{e}_j$ are unit vectors on the $j$-th direction.

The evolved reduced density matrix of the defects can now be obtained by assuming a thermal environment and integrating over the bath degrees of freedom. After a lengthy but straight forward calculation, the matrix elements in the eingenbasis $\{\ket{s_i}\}$ of the $S_x^{S,A}$ operator are given by
\begin{eqnarray}
&\bra{s_i}\rho_d(t)\ket{s_j}\!=\!\exp\!
\left\{\!-\!\big[f^S(t)(s_i^{S}\!-\!s_j^{S})^2\!+\!f^A(t)(s_i^{A}\!-\!s_j^{A})^2\big]\right. \nonumber \\
&\left.+i\big[\varphi^S(t)((s_i^{S})^2\!-\!(s_j^{S})^2)\!+\!\varphi^A(t)((s_i^{A})^2\!-(\!s_j^{A})^2)\big]\right\} \nonumber \\
&\times \bra{s_i}\rho_d(0)\ket{s_j},
%\label{eq:elements_rho}
\end{eqnarray}
where the coefficients $f^{S,A}(t)$ and $\varphi^{S,A}(t)$ are given by
\begin{eqnarray}
f^{S,A}(t)=&\sum_{i} \frac{\left(\widetilde\gamma_i^{S,A}\right)^2(2\widetilde n^{S,A}_i-1)}{2\left(\widetilde\omega_i^{(S,A)}\right)^3}\left(1-\cos\left(\widetilde\omega_i^{S,A}t\right)\right),
\label{eq:f_pm_app}\\
\varphi^{S,A}(t)=&\sum_{i} \frac{\left(\widetilde\gamma_i^{S,A}\right)^2}{2\left(\widetilde\omega_i^{(S,A)}\right)^2}\left(t-\frac{\sin\left(\widetilde\omega_i^{S,A}t\right)}{\widetilde\omega_i^{S,A}}\right).
\end{eqnarray}
and $\widetilde n^S_i$ ($\widetilde n^A_i$) is the thermal occupation number of mode $i$ of the symmetric (antisymmetric) bath defined by $\widetilde n^{S,A}_i=\displaystyle (e^{\widetilde\omega_i^{S,A}/T}-1)^{-1}$.

\bibliographystyle{unsrt}

\end{document}